\documentclass[twocolumn,floats,floatfix,showpacs,amssymb,prd,superscriptaddress,nofootinbib]{revtex4-1}
\usepackage{graphicx, epsfig, amssymb} 
\usepackage{epstopdf}
\usepackage{amsmath, amsfonts}
\usepackage{bm} 

\usepackage[linktocpage]{hyperref}
\usepackage[caption=false]{subfig}
\usepackage[usenames]{color}

\def\be{\begin{equation}}
\def\ee{\end{equation}}
\def\beq{\begin{eqnarray}}
\def\eeq{\end{eqnarray}}
\def\non{\nonumber}

\newcommand{\bea}{\begin{eqnarray}}
\newcommand{\eea}{\end{eqnarray}}
\newcommand{\ben}{\begin{enumerate}}
\newcommand{\een}{\end{enumerate}}
\newcommand{\bi}{\begin{itemize}}
\newcommand{\ei}{\end{itemize}}

\newcommand{\risco}{r_\mathrm{ISCO}}
\newcommand{\eps}{\epsilon}

\def\Schw{Schwarzschild }
\def\cN{{\cal N}}

\begin{document}

\title{\large Comparing numerical and analytical calculations of post-ISCO
  ringdown amplitudes}

\author{Shahar Hadar} \email{shaharhadar@phys.huji.ac.il}
\affiliation{Racah Institute of Physics, Hebrew
University,Jerusalem 91904, Israel.}

\author{Barak Kol} \email{barak\_kol@phys.huji.ac.il}
\affiliation{Racah Institute of Physics, Hebrew
University,Jerusalem 91904, Israel.}

\author{Emanuele Berti} \email{berti@phy.olemiss.edu}
\affiliation{Department of Physics and Astronomy, The University
of Mississippi, University, MS 38677, USA} \affiliation{California
Institute of Technology, Pasadena, CA 91109, USA}

\author{Vitor Cardoso} \email{vitor.cardoso@ist.utl.pt}
\affiliation{CENTRA, Departamento de F\'{\i}sica,
 Instituto Superior T\'ecnico, Universidade T\'ecnica de Lisboa - UTL,
 Av.~Rovisco Pais 1, 1049 Lisboa, Portugal.} \affiliation{Department of Physics and Astronomy, The
University of Mississippi, University, MS 38677, USA}

\date{\today}

\begin{abstract}
We numerically compute the ringdown amplitudes following the plunge of a
particle from the innermost stable circular orbit (ISCO) of a Schwarzschild
black hole in the extreme-mass ratio limit. We show that the ringdown
amplitudes computed in this way are in good agreement with a recent analytical
calculation \cite{Hadar:2009ip}.
\end{abstract}


\maketitle
\section{Introduction}

In this paper we will study a compact object plunging into a much more
massive, nonrotating (Schwarzschild) black hole from the innermost stable
circular orbit (ISCO), located (in \Schw coordinates) at $\risco=3 r_s$, where
$r_s=2M$ is the \Schw radius and $M$ is the black hole mass. This post-ISCO
plunge trajectory is of special interest because it is ``universal''. It has
long been known \cite{Peters} that the eccentricity of bodies orbiting a black
hole must decrease in the Newtonian regime of low velocities and large
separations ($r\gg r_s$) during a gravitational-wave driven inspiral . In this
sense, a plunge from a quasicircular ISCO represents a ``Keplerian
attractor'': for long, gravitational-wave driven inspirals the eccentricity
should be essentially zero by the time the particle reaches the ISCO. However,
some astrophysical scenarios do predict the possibility of orbits retaining
nonzero eccentricity all the way down to plunge (see e.g.
\cite{BarackCutler,HopmanAlexander,AmaroSeoane:2007aw,Yunes:2009yz}).

This ``universal'' plunge trajectory for particles falling from a
quasicircular ISCO leaves a very specific signature in the quasinormal ringing
of the final black hole. The amplitudes of the quasinormal modes excited in
the process were computed in Ref.~\cite{Hadar:2009ip}. In this paper we
confirm the predictions of that paper by computing gravitational radiation
with a frequency-domain perturbative code developed and tested in
Ref.~\cite{Berti:2010ce}, and we verify that the two calculations are in very
good agreement.
Therefore any model for extreme mass ratio inspirals leading to plunge from a
quasicircular ISCO should match the ringdown signal predicted in
Ref.~\cite{Hadar:2009ip} at late times. It will be interesting to generalize
the present results to comparable mass ratio binaries.

The paper is organized as follows. In section \ref{ppart} we describe the
post-ISCO plunge trajectory and the numerical algorithm to compute the
gravitational radiation produced by the plunging particle, along with the
resulting waveforms. In section \ref{sec:extract} we extract the ringdown
amplitudes from these waveforms and present a comparison with the results of
Ref.~\cite{Hadar:2009ip}.

\section{Radiation sourced by infalling object \label{ppart}}

The post-ISCO plunge trajectory is described by the coordinates $t,r,\phi$ as
a function of the proper time $\tau$ (without loss of generality we take the
orbit to lie in the equatorial plane, i.e. $\theta=\pi/2$).  The trajectory is
a solution of the geodesic equations
\bea \tilde{E} &=& f(r)\, \frac{dt}{d\tau}
\label{geodesic motion equations-tdot}\\
 \tilde{L} &=& r^{2}\, \frac{d\phi}{d\tau}
\label{geodesic motion equations-phidot}\\
 \tilde{E}^{2} &=& \left(\frac{dr}{d\tau}\right)^{2}+f(r)(\tilde{L}^{2}/r^{2}+1)
\label{geodesic motion equations-rdot} ~,
\eeq
where $f(r)=1-r_{s}/r$ and the energy and angular momentum per unit mass have
the values corresponding to a particle at the ISCO:
\bea\label{ELz}
\tilde{E} &=& \tilde{E}_\mathrm{ISCO}\equiv
\frac{2 \, \sqrt{2}}{3} \\\non \tilde{L} &=& \tilde{L}_\mathrm{ISCO} \equiv
\sqrt{3} \, r_{s} ~.
\eea
The plunge trajectory can be written down analytically: cf. Eq.~(2.7)--(2.11)
and Fig.~1 of Ref.~\cite{Hadar:2009ip}.
In perturbation theory, the gravitational radiation at infinity can be
determined from a knowledge of the Sasaki-Nakamura wave function $X_{lm}$ \cite{Sasaki:1981kj} (for
the extensive literature on gravitational waves from particles falling into
black holes, see \cite{Berti:2010ce,Nakamura:1987zz} and Appendix C of
\cite{Berti:2007fi}). In the frequency domain, the Sasaki-Nakamura equation
can be written in the form
\be
\frac{d^2X_{lm}}{dr_*^2} +
\left[
\omega^2-\frac{\Delta}{r^5}\left(l(l+1)r-6M\right)
\right]X_{lm}=S_{lm}\,.
\label{SNeq}
\ee
Here $(l,\,m)$ are (tensor) spherical harmonic indices resulting from a
separation of the angular variables, $\omega$ is the Fourier frequency of the
perturbation and $\Delta\equiv r(r-2M)$.  The boundary conditions dictate that
we should have outgoing waves at infinity and ingoing waves at the BH horizon:
\be
X_{lm}=\left\{
\begin{array}{l}
X_{lm}^{\rm in}e^{-i\omega r_*}\,,\quad r_*\to-\infty\,,\\
X_{lm}^{\rm out}e^{i\omega r_*}\,,\quad r_*\to+\infty\,.\\
\end{array}
\right.
\ee
The source term $S_{lm}$ in the Sasaki-Nakamura equation (\ref{SNeq}) is
determined by the point-particle trajectory, and it can be found in
Ref.~\cite{Berti:2010ce}. In terms of the Sasaki-Nakamura wavefunctions, the
plus- and cross- polarization amplitudes are given by
\beq
h_{+}+ih_{\times}&=&\sum_{lm} \,_{-2}Y_{lm} \left(h_{+\,lm}+ih_{\times\,lm}\right)\\
&=&\frac{8}{r}\int_{-\infty}^{+\infty}d\omega \sum_{lm}e^{i\omega(r_*-t)}\, _{-2}Y_{lm}X^{\rm out}_{lm}\,.
\eeq
In our comparisons we will always consider, for simplicity, the outgoing
amplitudes of the Sasaki-Nakamura wavefunction in the time domain, i.e.
\be
X^{\rm out}_{lm}(t)\equiv \int_{-\infty}^{+\infty}d\omega e^{i\omega(r_*-t)}X^{\rm out}_{lm}(\omega)\,.
\ee%

Numerically, the problem is to determine the (complex) amplitudes $X^{\rm
  out}_{lm}(t)$ for plunge trajectories with energy and angular momentum given
by Eq.~(\ref{ELz}).
In order to start the plunge, we must displace the particle from the ISCO
location by a small quantity $\epsilon$:
\be r_0= r_{\rm ISCO} \left(1-\epsilon \right) ~. \ee
We will present comparisons for $\epsilon=5\times 10^{-3}$ and
$\epsilon=10^{-2}$, but we verified that our results are robust by computing
the radiation for several other values of $\epsilon$, including $\epsilon=
5\times 10^{-4},\, 10^{-3},\, 5\times 10^{-3},\, 10^{-3},\, 5\times 10^{-3},\,
10^{-2}$.

\begin{figure}
\includegraphics[scale=0.3,clip=true]{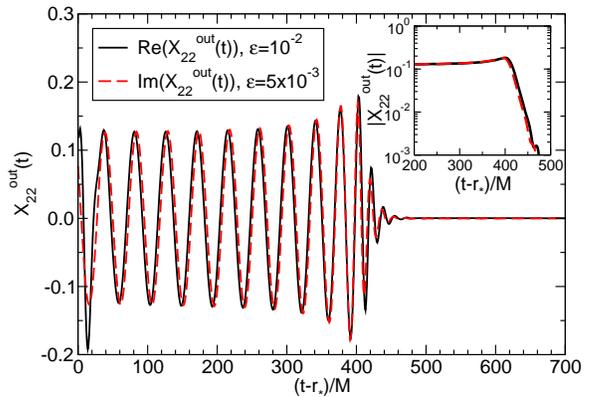}
\caption{\label{wfs} Visual comparison of two waveforms for $l=m=2$ with
  differing cutoffs ($\eps=10^{-2},\, 5\times 10^{-3}$). The time axis has been
  shifted so that the maxima of the wave amplitudes $|X_{lm}(t)|$ (shown in
  the inset) occur at the same value of $(t-r_*)/M$.}
\end{figure}

Our numerical integrations use a modification of the C++ program described in
Ref.~\cite{Berti:2010ce}, and we refer the reader to that paper for more
details. All differential equations are integrated in {\sc C++} using the
adaptive stepsize integrator {\sc StepperDopr5} \cite{NR}. First we integrate
the solution of the homogeneous SN equation (\ref{SNeq}) with ingoing boundary
conditions at the horizon from $r_h = 2M(1+\delta r)$ outwards (typically we
choose $\delta r=10^{-4}$). Then we integrate the solution with outgoing
boundary conditions at infinity from $r_{\infty}=r_{\infty}^{(0)}/\omega$
inwards, where typically we choose $r_{\infty}^{(0)} = 2\times 10^3$. From
these two independent solutions we compute the Wronskian at the (large but
finite) radius $r_{\infty}^{(0)}$. We integrate the geodesics with given
orbital parameters, and at the same time we compute the source term
corresponding to this trajectory. We output the solution with ingoing boundary
conditions at the horizon $X_{\rm in}^{(0)}$ and the source term $S_{lm}$ on
three numerical grids (each consisting of $n=1.6\times 10^5$ collocation
points). The three grids cover the intervals $[r_\infty,\,r_+]$, $[r_{\rm
    ISCO},\,r_{\rm ISCO}-0.1]$ and $[r_{\rm ISCO}-0.1,\,r_h]$,
respectively. This is necessary to keep good accuracy in the near-ISCO region,
where the particle spends a lot of time when $\epsilon$ is small. We use a
Gauss-Legendre spectral integrator \cite{NR} to compute the convolution of
$X_{\rm in}^{(0)}$ with the source term to find the outgoing wave amplitude
$X_{lm}^{\rm out}(\omega)$. Finally we sum over multipoles to get the total
radiated energy, angular momentum and linear momentum, and we perform a
Fourier transform to get $X_{lm}^{\rm out}(t)$.

The real part of the $l=m=2$ waveforms for
cutoff values $\epsilon=5\times
10^{-3}$ and $\epsilon=10^{-2}$ are shown in Figure \ref{wfs}. This visual
comparison shows quite clearly that the ringdown portion of the signal is very
weakly dependent on $\epsilon$.



\section{Extracting Ringdown Amplitudes from Waveforms \label{sec:extract}}
%
In this section we describe the procedure we used to extract the ringdown
amplitudes from the numerical waveforms and to compare with
Ref.~\cite{Hadar:2009ip}.
%
%
%
%

In the ringdown phase, the gravitational waveform is described by a
superposition of quasinormal modes of the form
\be X_{lm} = {\cN}
\sum_{n\, l\, m} R_{nlm}\, \exp i \omega_{nlm} (t-t_0)
\ee
where we dropped the ``out'' superscript for simplicity. Here $\omega_{n l m}$
are the (complex) characteristic ringdown frequencies, $R_{n l m}$ are the
amplitudes, $\cN$ is an overall normalization and $t_0$ is a time shift which
determines the origin of time.
We stress that while some readers may be more familiar with the odd and even
radiation functions $\psi_{lm}^{(odd)}$ (Regge-Wheeler) and
$\psi_{lm}^{(even)}$ (Zerilli), here we work with the Sasaki-Nakamura
radiation functions $X_{lm}$, which do not have a
specific parity.
%

Given a numerical time-domain waveform $X_{lm}(t)$, our objective is to
extract the amplitudes $R_{nlm}$. The leading amplitude ($n=1$) is extracted
by plotting $\log \left|X_{lm} \right|$ as a function of $t$ (see the inset of
Figure \ref{wfs}).  At late times only the dominant mode contributes to the
signal, and $\log \left|X_{lm} \right|$ becomes a linear function of time with
slope given by the decay constant $\gamma_1 \equiv \Im (\omega_{1lm})$. The
intercept is exactly the desired amplitude, $\log \left|R_{lm} \right|$. The
discrepancy between the expected and measured $\gamma_1$ contributes to the
error estimate for $|R_{1lm}|$. In practice, this contribution is minimized if
the origin of time is close to the center of the linear portion in the curve.

In the same fashion one can extract the amplitude of the second overtone,
namely $|R_{2lm}|$. In preparation for this, one should first extract the
phase of $R_{1lm}$. This is done by plotting $X'_{lm}=X_{lm} \cdot \exp
(\gamma_{1lm} t)$ as a function of $t$, so that $X' \simeq R \exp \left(i \Re
(\omega_{1lm}) t\right)$ should be a periodic function with frequency $\Re
(\omega_{1lm})$ and complex amplitude $R_{1lm}$.
Once we know the (complex) amplitude $R_{1lm}$, we can subtract the dominant
ringdown mode to obtain a residual $X^{(2)}_{lm}(t)= X_{lm}(t) - R_{1lm}\,
\exp i \omega_{1lm} t$. Now we can repeat the steps above to find the
amplitude of the second overtone. Indeed, the procedure can be repeated for
generic overtone numbers $n$ as long as the signal is not dominated by
numerical noise, i.e., as long as the residuals $X^{(n)}_{lm}$ exhibit a
decaying exponential behavior.

The value of the amplitudes depends, in principle, on the constants $\cN,
t_0$.
To sidestep this difficulty we compute the discrepancy of each mode
\be
\Delta \equiv \log \left(R^{\rm anlyt}_{nlm}/R^{\rm num}_{nlm}\right)\,,
\ee
where $R^{\rm anlyt}_{nlm}$ is the ``analytical'' amplitude computed in
Ref.~\cite{Hadar:2009ip} and $R^{\rm num}_{nlm}$ is the numerically extracted
amplitude. Next we plot $\Delta$ as a function of the mode decay constant
$\gamma_{nlm}$. Clearly a rescaling by $\cN$ will shift $\Delta$ by a
constant, while a shift of $t_0$ corresponds to a change in slope. In
particular, for modes with identical decay constants (such as modes with
different $m$ and fixed $l$) the discrepancy $\Delta$ is the same and it is
independent of $t_0$. Now we can carry out a linear fit of
$\Delta_{nlm}=\Delta_{nlm}(\gamma_{nl})$. This fit will yield the constants
$\cN, t_0$ required for the amplitudes to agree, i.e., $R^{\rm anlyt}_{nlm} =
R^{\rm num}_{nlm}$. The origin of time resulting from the linear fit was
compared to the conventions of Ref.~\cite{Hadar:2009ip}, where it what chosen
such that $r(t=0)=1.1 r_s$. We found that the same happens here to a good
approximation.

We performed comparisons for 16 different modes with the following $(nlm)$
values (for $\eps=0.005$): the modes $2 \le l \le 4$ for $0 \le m \le l$ and
$l = 5$ for $2 \le m \le 5$, all with $n=1$.

\begin{figure}[h!]
\includegraphics[scale=0.3,clip=true]{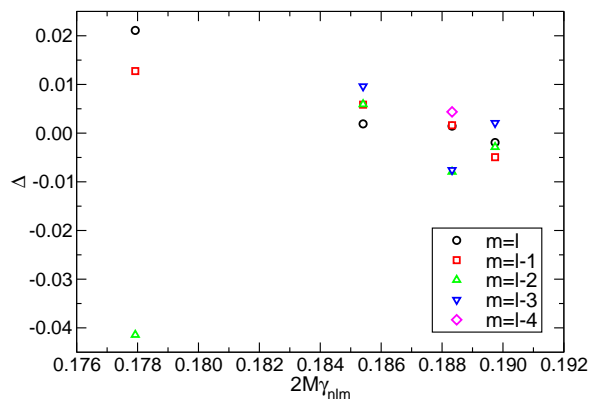}
\caption{\label{checkplot} Amplitude comparison for 16 different modes with
  $n=1$ and $\epsilon = 5\times 10^{-3}$. The horizontal axis represents the
  decay constant $\gamma_{nlm}$ in units where $r_{s} = 1$. Each point
  represents the residual $\Delta$ value (after subtracting a linear fit or
  tuning $\cN, t_{0}$) for the given $(l,\,m)$. Different $\gamma_{nlm}$
  values correspond to different $l$'s: the leftmost points correspond to
  $l=2$, and $\gamma_{nlm}$ grows monotonically with $l$.}
\label{residual error}
\end{figure}

Our main results are shown in Figure \ref{residual error}. After the linear
fit, the residual errors are seen to be of $\sim
1 \%$. We also
performed the comparison for the mode $(nlm) = (2 2 2)$ (second overtone of $l
= m = 2$). In this case we found good agreement up to a residual error of $\sim
10 \%$, consistent with our error estimate for the numerically extracted
amplitude in this case.  A similar comparison was made also for $\eps=0.01$
for some modes, and good agreement was found - up to an overall estimated error of $\sim
1 \%$.

In summary, for all the tested modes (see the list above) the numerical
amplitudes here were found to coincide with the analytic ones computed in
\cite{Hadar:2009ip}.


%
%

\noindent\textit{Acknowledgements:} This work was supported by the {\it
  DyBHo--256667} ERC Starting Grant, NSF PHY-090003 and FCT - Portugal through
PTDC projects FIS/098025/2008, FIS/098032/2008, CTE-AST/098034/2008, and
CERN/FP/109290/2009. E.B.'s research was supported by NSF Grant
No.~PHY-0900735.

\end{document}